\documentstyle[12pt,epsfig]{article}
\def\beq{\begin{equation}}
\def\eeq{\end{equation}}
\def\bea{\begin{eqnarray}}
\def\eea{\end{eqnarray}}

\begin{document}

\begin{center}
{\Large \bf \sf Entangled Quantum State Discrimination using Pseudo-Hermitian System. 
  }

\vspace{1.3cm}

{\sf Ananya Ghatak \footnote{e-mail address: \ \ gananya04@gmail.com}
and Bhabani Prasad Mandal \footnote{e-mail address:
\ \ bhabani.mandal@gmail.com, \ \ bhabani@bhu.ac.in  }}

\bigskip

{\em Department of Physics,\\
Banaras Hindu University,\\
Varanasi-221005, INDIA. \\
}

\bigskip
\bigskip

\noindent {\bf Abstract}

\end{center}

We demonstrate how to discriminate two non-orthogonal, entangled quantum state which are slightly different from each other by using pseudo-Hermitian system. The positive definite metric operator which makes the pseudo-Hermitian systems fully consistent quantum theory is used for such a state discrimination. We further show that non-orthogonal states can evolve through a suitably constructed pseudo-Hermitian Hamiltonian to orthogonal states. Such evolution ceases at exceptional points of the pseudo-Hermitian system.



\medskip
\vspace{1in}
\newpage
Suppose a particular quantum system is described by two states, $\mid\psi_{1}\rangle$ and $\mid\psi_{2}\rangle$ which are non-orthogonal $\langle\psi_{1}\mid\psi_{2}\rangle\not = 0$ and differ slightly ${\left|\langle\psi_{1}\mid\psi_{2}\rangle\right|}^{2}\cong 1-O(\epsilon^{2}), \epsilon\ll1$. At any instant of time the system will be either in state $\mid\psi_{1}\rangle$ or in state $\mid\psi_{2}\rangle$. However it is not possible to determine the state of such a system with a few measurements as $\mid\psi_{1}\rangle$ and $\mid\psi_{2}\rangle$ differ slightly. This problem of quantum state discrimination is very important in quantum information theory \cite{qit}. Recently Bender et. al. have proposed an alternative approach \cite{ben1} to discriminate two pure quantum states using the idea of PT-symmetric non-Hermitian quantum theory \cite{ben4}-\cite{cmba}. In PT-symmetric non-Hermitian theories one needs to construct a CPT-inner product \cite{cmba} to have a fully consistent quantum theory with unitary time evolution \cite{ben3}. Same inner product has been used to discriminate such states \cite{ben1}. However their discussion was restricted to two pure states. In this article we would like to extend the work in Ref. \cite{ben1} for two entangled quantum states using pseudo-Hermitian system \cite{am}-\cite{all}. We show that like PT-symmetric non-Hermitian system, pseudo-Hermitian quantum system can also be useful for discriminating entangled quantum states. Every pseudo-Hermitian system leads to a fully consistent quantum theory with unitary time evolution in a different Hilbert space endowed with a positive definite inner product \cite{spec,all}. In the same Hilbert space the states $\mid\psi_{1}\rangle$ and $\mid\psi_{2}\rangle$ are shown to become orthogonal.
In an alternative approach we show that the non-orthogonal states $\mid\psi_{1}\rangle$ and $\mid\psi_{2}\rangle$ evolve in time to become orthogonal to each other and hence are discriminated. However one needs to find an appropriate Hamiltonian which is responsible for such an evolution. In this work we show that a pseudo-Hermitian Hamiltonian suitably defined in a Hilbert space can serve the purpose. The states $\mid\psi_{1}\rangle$ and $\mid\psi_{2}\rangle$ have an unitary time evolution through such a pseudo-Hermitian Hamiltonian to become orthogonal to each other. Such an unitary time evolution not surprisingly breaks at the possible exceptional points \cite{kato}-\cite{zaf2} of the pseudo-Hermitian system.

We consider the following entangled states,
\begin{eqnarray}
\mid\psi_{1}\rangle &=& \frac {1}{\sqrt{2}} \cos\frac{\theta}{2}\left [ \mid 0,1/2\rangle +\mid 1,-1/2\rangle \right ] + \frac {1}{\sqrt{2}} \ \sin\frac{\theta}{2} \left[\mid 0,-1/2\rangle +\mid 1,1/2\rangle\right ]; \nonumber \\
 \mid\psi_{2}\rangle &=& \frac {1}{\sqrt{2}} \cos\frac{\theta +2\epsilon}{2} \left[\mid 0,1/2\rangle +\mid 1,-1/2\rangle\right] + \frac {1}{\sqrt{2}} \sin\frac{\theta +2\epsilon}{2} \left[\mid 0,-1/2\rangle +\mid 1,1/2\rangle\right], \nonumber \\
\end{eqnarray}
where $\epsilon$  is a very small quantity and ${\left|\langle\psi_{1}\mid\psi_{2}\rangle\right|}^{2}\cong 1-\epsilon^2$. We have used the notation $\mid n,\frac{1}{2} m_{s}\rangle$, with $n$ is the eigenvalue for the number operator $a^{\dagger}a$, i.e. $a^{\dagger}a\mid n,\frac{1}{2} m_{s}\rangle=n\mid n,\frac{1}{2} m_{s}\rangle$ and $m_{s}=\pm 1$ are the eigenvalues of the operator $ \sigma_{z}$, i.e. $\sigma_{z}\mid n,\frac{1}{2} m_{s}\rangle=m_{s}\mid n,\frac{1}{2} m_{s}\rangle$. For the sake of simplicity we take $\theta = \pi /2-\epsilon$, so the states can be written as,
\begin{eqnarray}
\mid\psi_{1}\rangle &= &\frac {1}{\sqrt{2}} \cos\frac{\pi -2\epsilon}{4}\left[\mid 0,1/2\rangle +\mid 1,-1/2\rangle\right] + \frac {1}{\sqrt{2}} \ \sin\frac{\pi -2\epsilon}{4}\left[\mid 0,-1/2\rangle +\mid 1,1/2\rangle\right]; \nonumber \\
\mid\psi_{2}\rangle &=&\frac {1}{\sqrt{2}} \cos\frac{\pi +2\epsilon}{4} \left[\mid 0,1/2\rangle +\mid 1,-1/2\rangle\right] + \frac {1}{\sqrt{2}} \ \sin\frac{\pi +2\epsilon}{4} \left[\mid 0,-1/2\rangle +\mid 1,1/2\rangle\right]. \nonumber \\
\end{eqnarray}
These two non-orthogonal states can be made orthogonal with help of the metric operator associated with the pseudo-Hermitian theory described by,
\begin{equation}
H=\mu\vec\sigma\cdot {\mathbf{B}}+\hbar\omega a^{\dagger}a+\rho \left(\sigma_{+}a-\sigma_{-}a^{\dagger}\right),
\end{equation}
where $\rho$ is an arbitrary real parameter \cite{bpm}.

This system describes a spin 1/2 particle in the external magnetic field $\mathbf{B}$ coupled to a simple harmonic oscillator through some non-Hermitian interaction. Here $\sigma_{i}$'s are Pauli spin matrices, $\sigma_{\pm}=1/2(\sigma_{x}\pm\iota\sigma_{y})$ are spin projection operators. $a, a^{\dagger}$ are usual creation and annihilation operators for the simple harmonic oscillator states. $a\mid n\rangle = \sqrt n\mid n-1\rangle; \ a^{\dagger}\mid n\rangle = \sqrt {n+1}\mid n+1\rangle$ where $\mid n\rangle$ represents the eigenvectors for simple harmonic oscillator. Without losing any essential features of the system we can choose the external magnetic field along z-direction, the Hamiltonian then changes to,
\begin{equation}
H=\frac{\varepsilon}{2}\sigma_{z}+\hbar\omega a^{\dagger}a+\rho(\sigma_{+}a-\sigma_{-}a^{\dagger}),
\label{y1}
\end{equation}  
with $\varepsilon =2\mu B_{z}$. 
It can be checked that this non-Hermitian $(H\not= H^{\dagger})$ Hamiltonian is pseudo-Hermitian with respect to two different operators parity (P) and $\sigma_{z}$ i.e., $H^{\dagger}=PHP^{-1}$ and $H^{\dagger}=\sigma_{z}H\sigma_{z}^{-1}$ \cite{bpm}.
The ground state of the system is $\mid 0,-1/2\rangle$ with energy eigenvalue $-\varepsilon/2$
\begin{equation}
H\mid 0,-1/2\rangle = -\varepsilon/2\mid 0,-1/2\rangle ,
\end{equation}
but, $\mid 0,1/2\rangle$ is not a eigenstate of this system. However the state $\mid 0,1/2\rangle$ along with $\mid 1,-1/2\rangle$ create an invariant subspace in the space of states as,
\begin{eqnarray}
H\mid 0, \ \ 1/2\rangle &=& \varepsilon /2\mid 0, 1/2\rangle -\rho\mid 1,-1/2\rangle ; \nonumber\\
H\mid 1,-1/2\rangle &=& \rho\mid 0,1/2\rangle +(\hbar\omega -\varepsilon /2)\mid 1,-1/2\rangle . \nonumber\\
\end{eqnarray}
Eigenvalues and eigenfunctions of the corresponding Hamiltonian matrix in this subspace are the eigenvalues and eigenfunctions for the first and second excited states of the system respectively.
A general invariant subspace is consist of $\mid n,1/2\rangle$ and $\mid n+1,-1/2\rangle$ and the Hamiltonian matrix corresponds to this invariant subspace is,
\begin{equation}
H_{n+1} = \left(\begin {array}{clcr}
\varepsilon /2+n\hbar\omega &  \rho\sqrt{n+1} \\
-\rho\sqrt{n+1}          &  -\varepsilon /2+(n+1)\hbar\omega \\
\end{array} \right).
\label{h1}
\end{equation}
The eigenvalues of the Hamiltonian matrix are given by,  
\begin{equation}
\lambda^{\pm}_{n+1}=\frac{1}{2}\left[(2n+1)\hbar \omega\pm\sqrt{(\hbar\omega -\varepsilon)^{2}-4\rho^{2}(n+1)} \ \right].
\end{equation}
These eigenvalues are real provided $(\hbar\omega -\varepsilon)\geq 2\rho\sqrt
{n+1}$. Putting $(\hbar\omega -\varepsilon)\sin{\alpha}=2\rho\sqrt{n+1}$ the eigenvectors are,
\begin{eqnarray}
\psi_{n+1}^{+} & = & \left(\begin {array}{clcr}
\sin {\alpha}/2  \\
\cos {\alpha}/2  \\
\end{array} \right) ; \nonumber \\
 \psi_{n+1}^{-} & = & \left(\begin {array}{clcr}
\cos {\alpha}/2  \\
\sin {\alpha}/2  \\
\end{array} \right).
\end{eqnarray}
The eigenvectors and eigenvalues of $H_{n+1}$ coalesce at $\alpha =\pi/2$, which is the exceptional point of the pseudo-Hermitian system described by the Hamiltonian in Eq. (\ref{y1}).
The positive definite metric operator for this system then is calculated using spectral method \cite{spec} as,
\begin{equation}
\eta = \mid \psi_{n+1}^{+}\rangle\langle\psi_{n+1}^{+}| \ + \mid\psi_{n+1}^{-}\rangle\langle\psi_{n+1}^{-}\mid ,
\end{equation} 
where $\psi_{n+1}^{\pm}$ are eigenvectors of $H_{n+1}^{\dagger}=H \left(\rho \rightarrow -\rho\right)$,
\begin{eqnarray}
\psi_{n+1}^{+} & = & \left(\begin {array}{clcr}
\cos {\alpha}/2  \\
-\sin {\alpha}/2  \\
\end{array} \right) ; \nonumber \\
 \psi_{n+1}^{-} & = & \left(\begin {array}{clcr}
-\sin {\alpha}/2  \\
 \cos {\alpha}/2  \\
\end{array} \right).
\end{eqnarray}
So, the metric operator is, 
\begin{equation}
\eta = \left(\begin{array}{clcr}
1 & -\sin\alpha  \\
-\sin\alpha  & \ \ \ 1 \\
\end{array}\right) ,
\label{eta}
\end{equation}
which can be expressed using $\sigma_{\pm}$ as, 
\begin{equation}
\eta = \bf {I}-\sigma_{+}\sin\alpha -\sigma_{-}\sin\alpha .
\label{eta2}
\end{equation}
Exactly same metric operator can also be calculated using the Ref. \cite{all}. Comparison of different methods for calculating $\eta$ is reported \cite{our}. The pseudo-Hermitian system (described by the $H_{n+1}$ given in Eq. (\ref{h1})) is a fully consistent quantum theory in the Hilbert space endowed with the metric in Eq. (\ref{eta}). The states $\mid\psi_{1}\rangle$ and $\mid\psi_{2}\rangle$ become orthogonal in the same Hilbert space, $\langle\psi_{1}\mid\psi_{2}\rangle_{\eta} = 0$ subjected to the condition $\sin\alpha =\cos\epsilon$.

The normalized eigenvectors $\langle\psi_{1}\mid$ and $\langle\psi_{2}\mid$ in the new Hilbert space endowed with metric operator $\eta$ given in Eq. (\ref{eta2}) are written as, 
\begin{eqnarray}
\langle\psi_{1}\mid &=& \frac{1}{\sqrt{2}\sin\epsilon}\left(\cos\frac{\pi-2\epsilon}{4}-\sin\frac{\pi-2\epsilon}{4}\cos\epsilon \right)\left[\langle 0,1/2\mid +\langle 1,-1/2\mid\right] \nonumber \\ 
&+&\frac{1}{\sqrt{2}\sin\epsilon}\left(\sin\frac{\pi-2\epsilon}{4}-\cos\frac{\pi-2\epsilon}{4}\cos\epsilon \right)\left[\langle 0,-1/2\mid +\langle 1,1/2\mid \right] ; \nonumber \\ 
\langle\psi_{2}\mid &=& \frac{1}{\sqrt{2}\sin\epsilon}\left(\sin\frac{\pi+2\epsilon}{4}-\cos\frac{\pi+2\epsilon}{4}\cos\epsilon\right)\left[\langle 0,1/2\mid +\langle 1,-1/2\mid\right] \nonumber \\
&+&\frac{1}{\sqrt{2}\sin\epsilon}\left(\cos\frac{\pi+2\epsilon}{4}-\sin\frac{\pi+2\epsilon}{4}\cos\epsilon \right)\left[\langle 0,-1/2\mid +\langle 1,1/2\mid\right] . \nonumber \\
\end{eqnarray}
However to construct the projection operator we need to consider all the states of the invariant subspace of the system described by $H_{n+1}$ in Eq. (\ref{h1}). In this case each invariant sub-Hilbert space consists of two states. Therefore we consider the other state as,
\begin{eqnarray}
\mid\psi_{3}\rangle &=&\frac {1}{\sqrt{2}} \sin\frac{\theta}{2} \left[\mid 0,1/2\rangle +\mid 1,-1/2\rangle\right] + \frac {1}{\sqrt{2}} \ \cos\frac{\theta}{2} \left[\mid 0,-1/2\rangle +\mid 1,1/2\rangle\right] . \nonumber \\
\end{eqnarray}
The non-orthogonal state which is slightly different from $\mid\psi_{3}\rangle$ is
\begin{eqnarray}
\mid\psi_{4}\rangle &=&\frac {1}{\sqrt{2}} \sin\frac{\theta-2\epsilon}{2} \left[\mid 0,1/2\rangle +\mid 1,-1/2\rangle\right] + \frac {1}{\sqrt{2}} \ \cos\frac{\theta-2\epsilon}{2} \left[\mid 0,-1/2\rangle +\mid 1,1/2\rangle\right] . \nonumber \\
\end{eqnarray}
It is interesting to see that these two non-orthogonal entangled states are also discriminated by the same metric operator $\eta$ in Eq. (\ref{eta}).  
Now we construct the projection operator $P=\Sigma P_{i}$, where $P_{i}$ are calculated as,
\begin{eqnarray}
P_{1}&=&\mid\psi_{1}\rangle\langle\psi_{1}\mid =\frac{1}{4 \sin\epsilon}\left[(1+\sin\epsilon )(A)+(\sin\epsilon -1)(B)-(\cos\epsilon )(C)+(\cos\epsilon )(D)\right] ; \nonumber \\
P_{2}&=&\mid\psi_{2}\rangle\langle\psi_{2}\mid =\frac{1}{4 \sin\epsilon}\left[(\sin\epsilon -1 )(A)+(\sin\epsilon +1)(B)+(\cos\epsilon )(C)-(\cos\epsilon )(D)\right] ; \nonumber \\
P_{3}&=&\mid\psi_{3}\rangle\langle\psi_{3}\mid =\frac{1}{4 \sin\epsilon}\left[(\sin\epsilon -1 )(A)+(\sin\epsilon +1)(B)+(\cos\epsilon )(C)-(\cos\epsilon )(D)\right] ; \nonumber \\
P_{4}&=&\mid\psi_{4}\rangle\langle\psi_{4}\mid =\frac{1}{4 \sin\epsilon}\left[(1+\sin\epsilon )(A)+(\sin\epsilon -1)(B)-(\cos\epsilon )(C)+(\cos\epsilon )(D)\right] . \nonumber \\
\end{eqnarray} 
where A, B, C, D are denoted as below,
\begin{eqnarray}
A =\mid 0,\frac{1}{2}\rangle\langle 0,\frac{1}{2}\mid +\mid 0,\frac{1}{2}\rangle\langle 1,-\frac{1}{2}\mid +\mid 1,-\frac{1}{2}\rangle\langle 0,\frac{1}{2}\mid +\mid 1,-\frac{1}{2}\rangle\langle 1,-\frac{1}{2}\mid ; \nonumber \\
B =\mid 0,-\frac{1}{2}\rangle\langle 0,-\frac{1}{2}\mid +\mid 0,-\frac{1}{2}\rangle\langle 1,\frac{1}{2}\mid +\mid 1,\frac{1}{2}\rangle\langle 0,-\frac{1}{2}\mid +\mid 1,\frac{1}{2}\rangle\langle 1,\frac{1}{2}\mid ; \nonumber \\
C =\mid 0,\frac{1}{2}\rangle\langle 0,-\frac{1}{2}\mid +\mid 0,\frac{1}{2}\rangle\langle 1,\frac{1}{2}\mid +\mid 1,-\frac{1}{2}\rangle\langle 0,-\frac{1}{2}\mid +\mid 1,-\frac{1}{2}\rangle\langle 1,\frac{1}{2}\mid ; \nonumber \\
D =\mid 0,-\frac{1}{2}\rangle\langle 0,\frac{1}{2}\mid +\mid 0,-\frac{1}{2}\rangle\langle 1,-\frac{1}{2}\mid +\mid 1,\frac{1}{2}\rangle\langle 0,\frac{1}{2}\mid +\mid 1, \frac{1}{2}\rangle\langle 1,-\frac{1}{2}\mid . \nonumber \\
\end{eqnarray}
It is straight forward to check that $ P=\Sigma_i P_i=\bf {I} $. The projection operator $P_{1}$ (or $P_{2}$) can be used to discriminate the states $\mid\psi_{1}\rangle$ and $\mid\psi_{2}\rangle$. On the other hand application of either $P_{3}$ or $P_{4}$ discriminates $\mid\psi_{3}\rangle$ from $\mid\psi_{4}\rangle$. $P_{i}$ projects the states $\mid\psi_{i}\rangle$ for $i=1,2,3,4$.
\\ \\
Alternatively these states can be discriminated through time evolution. Two non-orthogonal states which differ slightly at some initial time can evolve through suitably chosen pseudo-Hermitian Hamiltonian to two orthogonal states at later time. We start with non-orthogonal states $(\psi_{1},\psi_{2})$ or $(\psi_{3},\psi_{4})$ at $t=0$ i.e. $\langle\psi_{1}(t=0)\mid\psi_{2}(t=0)\rangle\not =0\not =\langle\psi_{3}(t=0)\mid\psi_{4}(t= 0)\rangle$. Here standard Dirac inner product rule i.e. complex conjugate and transpose is used. The pseudo-Hermitian Hamiltonian H in Eq. (\ref{y1}) can be used to discriminate the states $\mid\psi_{1}\rangle$ and $\mid\psi_{2}\rangle$ at some later time. The states $\mid\psi_{1}\rangle$, $\mid\psi_{2}\rangle$ evolve through the pseudo-Hermitian H in Eq. (\ref{y1}) to two orthogonal states.
To show this we write the Hamiltonian in Eq. (\ref{y1}) as, 
\begin{equation}
H=\frac{1}{2}\hbar\omega {\mathbf{I}}+\vec{\sigma} \cdot \left[0, \ i\rho, \ \frac{1}{2}(\epsilon -\hbar\omega)\right] ,
\end{equation}  
and the Hermitian conjugate of the above Hamiltonian as, 
\begin{equation}
H^{\dagger}=\frac{1}{2}\hbar\omega {\mathbf{I}}+\vec{\sigma} .\left[0, \ -i\rho, \ \frac{1}{2}(\epsilon -\hbar\omega)\right] .
\end{equation}
Now to calculate that time evolution  $\langle\psi_{1}(t=0)\mid e^{iH^{\dagger}t}e^{-iHt}\mid\psi_{2}(t=0)\rangle$ we consider the term
\begin{eqnarray}
\cos^{2}\alpha \ e^{iH^{\dagger}t}e^{-iHt} \nonumber \\ 
=&\left(\begin {array}{clcr}
\cos^{2}\beta  t\cos^{2}\alpha +\sin^{2}\beta  t(1+\sin^{2}\alpha) \ \ \  &  \sin 2\beta  t\sin\alpha (-i\cos\alpha -\sin\beta  t) \\
\\
\sin 2\beta  t\sin\alpha (i\cos\alpha -\sin\beta  t)  \ \ \ \ \ &  \cos^{2}\beta  t\cos^{2}\alpha +\sin^{2}\beta  t(1+\sin^{2}\alpha) \\
\end{array} \right)& , \nonumber \\
\end{eqnarray}
where, $\beta =\sqrt{-\rho^{2}+\frac{1}{4}(\epsilon -\hbar\omega )^{2}}$ and the identity $e^{(i\phi\sigma\cdot n)}=\cos\phi \ {\mathbf{I}}+i\sin\phi \ \vec{\sigma}\cdot\vec{n}$ has been used. Now $\langle\psi_{1} (t)\mid\psi_{2} (t)\rangle =\langle\psi_{1}(t=0)\mid e^{iH^{\dagger}t}e^{-iHt}\mid\psi_{2} (t=0)\rangle=0$ if  
\begin{eqnarray}
\sin^2{\beta t}&=&\left(-4\cos\alpha\left(\left(1-3\cos 2\alpha \right)\cos\epsilon\cot\alpha +\left(1-\cos 2\epsilon -4\sin\alpha\right)\cos\alpha\right) \right.\nonumber \\ 
&+&\left(-4\cos^4\alpha \sin^2\alpha \left(\cos\epsilon -\sin\alpha\right)^2\left(\left(\cos 2\alpha +\cos\epsilon\sin\right)^2-4\cos^2\alpha\sin\alpha\right) \right.\nonumber \\ 
&+&\left.\left. \frac{1}{16}\left[\left(\cos 2\epsilon+4\sin\alpha -1\right)\sin^2 2\alpha +2\cos^2\alpha\cos\epsilon\left(3\sin 3\alpha -5\sin\alpha\right)\right]^2\right)^\frac{1}{2}\right) \nonumber \\ 
&.&\left(2\left(\left(\cos 2\alpha +\cos\epsilon\sin\alpha\right)^2 -4\cos^2 \alpha\sin\alpha\right)\right)^{-1}.  \nonumber \\ 
\end{eqnarray}
This equation has a definite solution for t except at $\alpha =\pi /2$ [Fig.1], which is the exceptional point in the spectrum of the pseudo-Hermitian system used for the time evolution.
\ \ \ 
 
\includegraphics{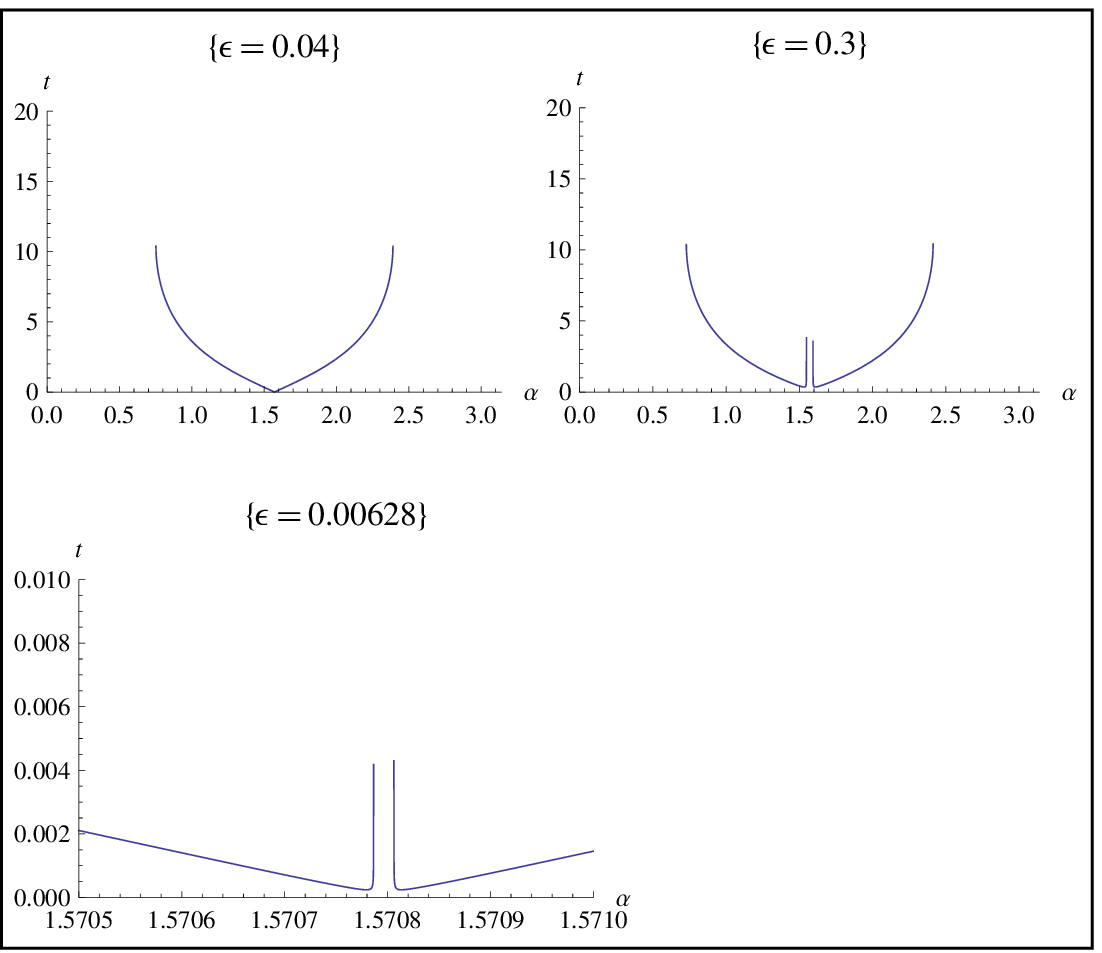}

Fig.1: The time-evolution of the non-orthogonal states with respect to the parameter in the pseudo-Hermitian system for different values of the small parameter $\epsilon$. The time evolution breaks down at $\alpha=\pi/2$, which is the exceptional points of the pseudo-Hermitian theory. \\

Conclusions: Discriminating two non-orthogonal states play very important role in quantum information theory. We have shown that two non-orthogonal entangled quantum states which are slightly different from each other can be discriminated by using suitably constructed pseudo-Hermitian systems. Pseudo-Hermitian system becomes fully consistent quantum theory in a different Hilbert space endowed with a positive definite metric. The same metric operator has been used to discriminate the non-orthogonal entangled states. Alternatively, we have shown that quantum entangled states which are non-orthogonal at $t=0$, can evolve through suitably chosen pseudo-Hermitian Hamiltonian to orthogonal states. Such time evolutions are obstructed by the existence of exceptional points of the pseudo-Hermitian system. We have demonstrated all these by considering explicit example.


\vspace{.2in}

\vspace{.2in}
                                                                                

                                                                             
\begin{thebibliography}{99} 
\bibitem{qit} A. Chefles {\em Lect. Notes Phys.} {\bf 649}, 467 Springer (2004).

\bibitem{ben1} C. M. Bender, D. C. Brody, J. Caldeira, B. K. Meister, {\em arXiv} {\bf 1011.1871} (2010).

\bibitem{ben4} C. M. Bender and S. Boettcher {\em Phys.Rev.Lett.} {\bf 80}, 5243 (1998).

\bibitem{ben2} C. M. Bender, {\em Rep.Prog. Phys.} {\bf 70} 947 (2007) and references therein.

\bibitem{ben3} C. M. Bender, D. C. Brody and H. F. Jones {\em Phys. Rev. Lett.} {\bf 89} 270401(2002) ; Erratum-ibid. {\bf 92} 119902(2004). 

\bibitem{new3} A. Khare and B. P. Mandal, {\em Spl issue of Pramana J of Physics} {\bf 73}, 387 (2009).

\bibitem{bm} B. Basu-Mallick, {\em Int. J. of Mod. Phys. B}  {\bf 16}, 1875 (2002); B. Basu-Mallick, T. Bhattacharyya, A. Kundu, and B. P. Mandal {\em Czech. J. Phys } {\bf 54}, 5 (2004).

\bibitem{sca1} B. Basu-Mallick and B.P. Mandal, {\em Phys. Lett.} {\bf A 284} 231 (2001); B. Basu-Mallick, T. Bhattacharyya  and B. P. Mandal, {\em
 Mod. Phys.Lett.} {\bf A 20 }, 543 (2004).
 
\bibitem {bsg} B. P. Mandal, S. Gupta, {\em Mod.Phys.Lett. A} {\bf 25} 1723 (2010).

\bibitem{cmba} C. M. Bender, B. Tan {\em J.Phys.} {\bf A39}, 1945, (2006); C. M. Bender, J. Brod, A. Refig, M. Reuter {\em J. Phys. A: Math. Gen.} {\bf 37} (2004) 10139; C. M. Bender, H. F. Jones, {\em Phys.Lett.A} {\bf 328}, 102 (2004).

\bibitem{am} A. Mostafazadeh, {\em Int. J. Geom. Meth. Mod. Phys.} {\bf 7}, 1191(2010) and references therein.

\bibitem{bpm} B. P. Mandal, {\em Mod. Phys. Lett.} {\bf A 20} 655(2005).

\bibitem{spec} A. Mostafazadeh and A. Batal, {\em J. Phys A: Math. and theor.} {\bf 37},11645(2004).

\bibitem{all} A. Das, L. Greenwood, {\em Phys. Lett. B} {\bf 678 5} 504 (2009); A. Das, L. Greenwood, {\em J.Math.Phys.} {\bf 51}, 042103, (2010).

\bibitem{zhm} Z. H. Musslimani, {\em Phys.Rev. Lett.}{\bf 100} 030402 (2008).

\bibitem{nat1} C. E. Ruter, K. G. Makris, R. El-Ganainy, D. N. Christodoulides, M. Segev, D. Kip, {\em Nature Physics} {\bf 6}, 192 (2010); Z. H. Musslimani, K. G. Makris, R. El-Ganainy, and D. N. Christodoulides, {\em Phys. Rev. Lett.} {\bf 100}, 030402 (2008).

\bibitem{ctt} C. T. West, T. Kottos, and T. Prosen, {\em Phys. Rev. Lett.} {\bf 104}, 054102 (2010).

\bibitem{guo} A. Guo, G. J. Salamo, D. Duchesne, R. Morandotti, M. Volatier-Ravat, V. Aimez, G.A. Siviloglou, and D.N. Christodoulides, {\em Phys. Rev. Lett.} {\bf 103}, 093902 (2009).

\bibitem{rut} C. E. Ruter, K. G. Makris, R. El-Ganainy, D. N. Christodoulides, M. Segev, and D. Kip, {\em Nature (London)Phys.} {\bf 6}, 192-195 (2010).

\bibitem{kato} T. Kato, {\em Perturbation Theroy of Linear Operators}, {\bf Springer}, Berlin, (1966).

\bibitem{bery} M.V. Berry, {\em Czech. J. Phys.} {\bf 54}, 1039 (2004).

\bibitem{hess} W. D. Heiss, {\em Phys. Rep.} {\bf 242}, 443 (1994).

\bibitem {gun} S. Klaiman, U. Gunther, and N. Moiseyev, {\em Phys. Rev. Lett.} {\bf 101}, 080402 (2008); M. M¨uller and I. Rotter, {\em J. Phys. A} {\bf 41}, 244 018 (2008).

\bibitem{amos} A. Mostafazadeh {\em Phys. Rev. Lett.} {\bf 102}, 220402 (2009).

\bibitem{slp} S. Longhi {\em Phys. Rev. B}  {\bf 80}, 165125 (2009).

\bibitem{bfs} B. F. Samsonov {\em J. Phys. A} {\bf 43}, 402006 (2010); B. F. Samsonov  {\em Math. J. Phys. A}: {\em Math. Gen.} {\bf 38}, L571 (2005).

\bibitem{zaf1} Z. Ahmed {\em arXiv:} {\bf 0908.2876} (2009).

\bibitem{zaf2} Z. Ahmed {\em J. Phys. A: Math. Theor.} {\bf 45}, 032004 (2012). 

\bibitem{our} A. Ghatak, B. P. Mandal, submitted in journal.


\end{thebibliography}
\end{document}